# Closing the Loop on Morphogenesis:
# A Mathematical Model of Morphogenesis by Closed-Loop Reaction-Diffusion

Joel Grodstein[1] and Michael Levin[2,3,*]

[1] Department of Electrical and Computer Engineering, Tufts University, Medford, Massachusetts, USA.
[2] Allen Discovery Center at Tufts University, Medford, Massachusetts, USA.
[3] Wyss Institute for Biologically Inspired Engineering at Harvard University, Boston, Massachusetts, USA.

\* Corresponding author

**Abstract**

Morphogenesis, the establishment and repair of emergent complex anatomy by groups of cells, is a fascinating and biomedically-relevant problem. One of its most fascinating aspects is that a developing embryo can reliably recover from disturbances, such as splitting into twins. While this reliability implies some type of goal-seeking error minimization over a morphogenic field, there are many gaps with respect to detailed, constructive models of such a process being used to implement the collective intelligence of cellular swarms. We describe a closed-loop negative-feedback system for creating reaction-diffusion (RD) patterns with high reliability. It uses a cellular automaton to characterize a morphogen pattern, then compares it to a goal and adjusts accordingly, providing a framework for modeling anatomical homeostasis and robust generation of target morphologies. Specifically, we create a RD pattern with *N* repetitions, where *N* is easily changeable. Furthermore, the individual repetitions of the RD pattern can be easily stretched or shrunk under genetic control to create, e.g., some morphological features larger than others. Finally, the cellular automaton uses a *computation wave* that scans the morphogen pattern unidirectionally to characterize the features that the negative feedback then controls. By taking advantage of a prior process asymmetrically establishing planar polarity (e.g., head vs. tail), our automaton is greatly simplified. This work contributes to the exciting effort of understanding design principles of morphological computation, which can be used to understand evolved developmental mechanisms, manipulate them in regenerative medicine settings, or embed a degree of synthetic intelligence into novel bioengineered constructs.

**Introduction: reaction/diffusion, positional information, and scaling**

The generation of complex form during embryonic development, and its repair and remodeling during regeneration, highlight fundamental problems that range from cell and evolutionary biology to control theory and basal cognition [1-4]. How can collections of cells cooperate to reliably produce the same species-specific target morphology? Moreover, what mechanisms enable them to robustly do so despite various perturbations? For example, planarian flatworms regenerate their entire body from large or small fragments of any type [5], while amphibian embryos maintain the right proportions even when many cells are missing [6, 7] or made too large [8, 9]. This homeostatic property of multicellular morphogenesis has numerous implications beyond basic science, as it represents an attractive target for regenerative medicine and synthetic bioengineering approaches that seek efficient methods for the control of growth and form.

Moreover, it has been analyzed as an example of collective intelligence, showing how groups of cells (competent in physiological and metabolic spaces) can solve problems in anatomical morphospace [10, 11]. One such problem is reliably organizing positional information de novo, for example along an embryonic axis. A number of mathematical frameworks have been developed to help understand, predict, and control the decision-making of cells in the morphogenetic problem space. Here, we first review several popular approaches to modeling this process, highlighting their positive features and limitations. We then propose a new model which has interesting and useful features for quantitative modeling of morphogenesis.

Reaction-diffusion (RD) and positional information (PI) are perhaps the two best known hypotheses in the field of morphogenesis. Green [12] gives an excellent summary of both hypotheses as well as contrasting the two. RD [12-18] was proposed by Turing in 1952. In its simplest form, it uses two chemical species, *A* and *I*. *A* (the "activator") generates more of *A* and/or *I* via chemical reactions; *I* (the "inhibitor") similarly reduces their concentrations. Surprisingly, combining these reactions with the diffusion of both *A* and *I* can, in many cases, amplify small random concentration gradients into definite and striking patterns (see [15] for many examples).

Intuitively, the activator *A* promotes more of both *A* and *I*. Thus, any small excess of *A* at any location quickly grows by positive feedback. Of course, [*I*] also grows at the same location; but *I* is assumed to diffuse faster than *A*, so there is soon relatively little of *I* at this peak, and so the peak stays a peak. The *I* near the peak prevents new peaks from forming until you get far enough away for [*I*] to drop, at which point the pattern repeats. This concept, local self-activation with long-range inhibition, has been the basis of most RD systems (though new versions have also been discovered[19, 20]). All of the variants have the basic ability to start with small, random concentration variations and amplify them into stable large-scale patterns.

Almost 20 years after Turing, Lewis Wolpert published his positional-information hypothesis [21]. It is attractively simple. First, some unspecified process creates a gradient of a morphogen from, say, head to tail. Next, cells use a gene regulatory network (GRN) to determine their position by sampling the morphogen gradient, and then differentiate accordingly. PI gained rapid popularity. But it never per se explained where the initial gradient came from. Furthermore, most morphogen gradients exhibit exponential decay, which implies that much of the field will contain very low concentrations. This would make it difficult[22] for a GRN to determine spatial locations in those areas.

RD is not a general solution, either. RD patterns have a characteristic length $\lambda_{RD}$, typically given by $\lambda_{RD} = \sqrt{\frac{D_A}{K_{D,A}}}$ (where $D_A$ is the diffusion constant of *A* in m$^2$/sec and $K_{D,A}$ is the degradation constant of $A$ in sec$^{-1}$). In a field of length $L$, an RD pattern typically repeats $L/\lambda_{RD}$ times. Thus, longer fields typically result in more pattern repetitions. This was originally seen as an argument against RD[16]; while it is reasonable for larger animals to have, say, more spots, we would not expect a larger embryo to have extra fingers or toes. This objection was eventually partially overcome. Gierer and Meinhardt first proposed [14] a scale-independent version of RD. The advent of modern molecular-biology techniques produced evidence[23] that mouse digits are formed with an RD system that uses feed-forward techniques – a molecule that affects embryo size also feeds forwards to affect $\lambda_{RD}$, thus keeping $\lambda_{RD}$ reasonably aligned with embryo size and tending to produce the correct number of digits.

Barkai later proposed *expansion-repression*[24, 25], which uses a morphogen $A$ and "expander" species $E$. $A$ is generated at one end of the field at $x=0$, and then diffuses and decays everywhere, again with a characteristic length of $\lambda_{RD} = \sqrt{\frac{D_A}{K_{D,A}}}$. Thus [$A$] falls off as $x > \lambda_{RD}$. Because $E$ is generated only when [$A$] is less than some repression threshold $T_{rep}$, then $E$ serves as a way to detect that $\lambda_{RD}$ is shorter than $L$. They propose that $E$ diffuses very quickly and causes $\lambda_{RD}$ to increase everywhere (either by increasing $D_A$ or by decreasing $K_{D,A}$). By using [$E$] to alter $\lambda_{RD}$, they robustly set $\lambda_{RD}=L/2$ and create exactly one repetition of an RD pattern, proportionally scaled to $L$ (Figure 1). This is a negative-feedback system, where [$E$] essentially measures $L/\lambda_{RD}$ and then feeds back to adjust $\lambda_{RD}$.

RD and PI were typically seen as competing theories, at least until experimental evidence mounted for each being used in different circumstances. Green [12] even suggested that RD and PI can work together in the same system, e.g., by having RD lay down a gradient that PI then uses. Expansion-repression is an example of this; it lays down a scalable one-peak pattern, thus creating a morphogen gradient varying from low at one end of an organism to high at the other. Since it is scale independent, the coordinate system scales with the length of the organism.

The central problem is how to take an existing set of features, which may or may not be correct, and move in the correct direction in morphospace. The capability of RD hypotheses to adapt to the field length $L$ has clearly improved over time; from not at all in Turing's original work[13], to the simple feed-forward hypothesis for mouse digits [23], to negative feedback in expansion-repression[24].

We instead use a cellular automaton as part of a closed-loop negative-feedback system. Instead of measuring a proxy for $L$, we will directly measure the number of RD pattern repetitions using the automaton, compare that to the target shape and iteratively adjust $\lambda_{RD}$ accordingly, all under digital control. Cellular automata have been used to navigate morphospace; e.g., by Mordvintsev[26] to robustly create images. However, [26] uses an entire neural network in each cell, which is an unrealistic amount of compute power. Our approach, by contrast, takes advantage of having already broken symmetries (e.g., distinguishing an animal's head from its tail) to greatly simplify the problem.

*Overview of our proposed closed-loop RD system using a cellular automaton*

Consider a simple RD pattern on a 1D field of cells. The number of replications of the basic RD pattern will depend on the length of the field and the pattern's intrinsic length $\lambda_{RD}$. Expansion-repression, as noted above, can alter $\lambda_{RD}$ so that we always get exactly one peak at the source. We go a step further: given an integer goal $N$, we will alter $\lambda_{RD}$ so as to obtain exactly $N$ peaks (Figure 1). To do this, we use a cellular automaton – an identical simple GRN in each cell – that serves to count the number of peaks. We then wrap the automaton in a top-level control loop that iteratively adjusts $\lambda_{RD}$ until we have exactly *N* peaks. Essentially, we have built a closed-loop negative-feedback goal-seeking machine for morphogenesis. It knows its target pattern shape and adjusts parameters iteratively until that goal is achieved.

We add one more capability. Green [12] has proposed systems where RD acts downstream of PI, with a morphogen gradient inducing a gradual increase in $\lambda_{RD}$ so that, e.g., digits in a mouse paw are wider at their distal end than their proximal end[27]. Meinhardt[17] has proposed a similar mechanism in Hydra. Both of these serve to build an RD pattern where $\lambda_{RD}$ varies from a small, tight pattern at one end of the field to a larger $\lambda_{RD}$ at the other end. As our automaton counts the peaks in an RD pattern, it leaves behind digital breadcrumbs such that each cell knows its exact ordinal position. A small amount of per-cell logic can then examine those signals and increase or decrease $\lambda_{RD}$ in any given cell(s). As a result (Figure 2), we too can make $\lambda_{RD}$ larger or smaller at different locations in the field; but we can do it arbitrarily, rather than only a simple monotonic increase from one end to the other.

**Materials and Methods: The cellular automaton**

In this section, we describe our cellular automaton for counting RD pattern peaks. A human, looking at the patterns in Figure 1, can easily tell that Figure 1a has one peak and Figure 1b has three. We would basically scan the picture from left to right, counting each peak as it occurs. But how can an organism, using a simple GRN replicated in every cell, perform this task? The "left to right" part of our human algorithm somehow needs to translate into practical biology. We start by assuming that we have a collection of different signaling molecules that, by some magic, diffuse directionally; e.g., only travel from left to right and, furthermore, travel only to the neighboring cell before decaying. We might use a set of signals *S0*, *S1*, *S2*, etc., denoting the number of peaks to any cell's left. Figure 3a shows which signals should be expressed in which locations as per this scheme.

Consider the following logic in each cell:
```
edge = ([A]_left<.3) and([A]_me>.3)
S0_out = S0_in and (not edge)
S1_out = (S0_in and edge) or (S1_in and (not edge))
S2_out = (S1_in and edge) or (S2_in and (not edge))
S2_out = (S1_in and edge) or (S2_in and (not edge)), etc.
```
By the definition of a cellular automaton, each cell must (and does) implement exactly the same logic. Furthermore, while we expressed this logic in terms of simple Boolean AND, OR and NOT gates, it can easily be translated into a GRN[28]. Finally, assume a top-level controller forcing the leftmost cell to express *S0*.

This implementation has several issues, of course. First, magical one-way signals are not realistic. Second, the GRN in each cell is looking for peaks, and does so by finding rising edges, and uses the [*A*] in the cell to its left to do so; but we've not specified how that concentration could be communicated across cells. Third, our automaton is not robust to noisy signals. For example, if [*A*] had some noise that caused wiggling around the cell where [*A*]=.3, our GRN might miscount the noise as an extra peak. We can resolve the second and third issues with a simple trick that is very common in human-designed noise-resistant filtering schemes, called a *Schmidt trigger*. Instead of simply having one signal to count each peak, we now use two; one for each low (L) and one for each high (H). Our new signals *S0L*, *S0H*, *S1L*, *S1H*, etc., are now generated as per Figure 3b.

Once the automaton detects a high signal (e.g., $[A]>.3$), it will not count the next low-going edge until the signal is fairly low (e.g., $[A]<.1$). At that point, it will not declare a new high-going edge until the signal is again $>.3$. This not only gives us excellent noise immunity (we are now nearly immune to double counting at noisy locations), but also removes the necessity of communicating $[A]$ between neighboring cells; each cell need only look at its own $[A]$. Here is the new automaton in each cell (where the top-level controller now seeds cell #0 with *S0L*):

```
very_low = ([A]me<.1)
very_high = ([A]me>.3)
S0L_out = S0L_in and (not very_high)
S0H_out = (S0L_in and very_high) or (S0H_in and (not very_low))
S1L_out = (S0H_in and very_low) or (S1L_in and (not very_high))
S1H_out = (S1L_in and very_high) or (S1H_in and (not very_low)), etc.
```

**Results**

Implementing unidirectional nearest neighbor signaling.

We are still left with the problem of how to implement a signaling species that only travels unidirectionally to its nearest neighbor cell. Real molecules, be they proteins or small molecules, diffuse bidirectionally and often over longer distances. The use of real-world signals would quickly break our automaton, as shown in Figure 4. *S0H* would not only diffuse to the right (as desired) but also to the left. When it reached a cell with $[A]<.1$, that cell would then generate *S1L*. *S1L* would then diffuse to the right, eventually creating *S1H* at the same cell that originally generated *S0H*. Loops such as this would quickly generate incorrectly-large counts.

We implement unidirectional signaling with what we call a *computation wavefront* [29-31]. Cell #0 (at the left) generates *S0L*. When cell #1 sees *S0L*, it notices that its $[A]$ is still too low to emit *S0H*, and instead decides to regenerate *S0L*. Once it makes this decision, it closes itself off to future decision making. Even if at some point later it eventually sees *S0H* (e.g., from a loop as above), it will not generate *S1L* because its decision has already been made. In other words, it interprets the first signal it sees as the approach of the computation wavefront, acts on that wavefront by sending out its own signal, and then will not change the signal it sends out until the system is reset (which happens before each new count).

With this system, even though our actual signaling molecules diffuse bidirectionally, the computation wavefront proceeds only from left to right. The key is that any path from cell #0 to a cell *C* via a loop must be longer than the simple direct path from cell #0 to *C*, and thus arrives at *C after* the direct signal has arrived, and thus cannot affect the action that cell *C* takes. The combination of Schmidt triggers with computation wavefronts counts peaks quite robustly; we will discuss the limits of automaton robustness shortly.

The top-level controller

The next piece in our system is the top-level controller. The controller takes a goal *N*. It uses the cellular automaton to count peaks; then compares the count to *N* and adjusts $\lambda_{RD}$ as needed; and iterates this sequence until we have exactly *N* peaks. It is conceptually quite simple:

```
start with initial λRD and wait for the pattern to settle
Loop:
reset the automaton and count the number of peaks
if (number of peaks == N) we're done
```

```
    else if (number of peaks is too large):
        increase λRD
    else # number of peaks is too small
        decrease λRD
        re-seed the pattern          # explained below
    go to Loop
```

What does the "re-seed the pattern" step mean? It has been known since the 1970s [32] that the number of repetitions of a Turing pattern is sensitive to initial conditions. Werner has noted [33] that while $L/\lambda_{RD}$ is an upper bound for the number pattern replications in a field of length $L$, the lower bound can sometimes be 1 (Figure 5). In other words, while we cannot fit (e.g.,) four RD pattern repetitions in a space only large enough for three, it is possible (albeit unlikely[33]) for one single pattern repetition to stretch/scale itself up to an almost arbitrarily large field.

The top-level code above thus includes a small trick. When we are *increasing* $\lambda_{RD}$, we merely continue the simulation with the larger value. But when we decrease $\lambda_{RD}$, this may not succeed at creating extra peaks. Instead, it turns out that setting $[I]=0$, with $[A]$ rising linearly from 0 at tail to 1 at the head is reasonably reliable at seeding the maximum number of peaks in a given field size. As discussed below, it is not 100% reliable – but our closed-loop controller successfully works around the unreliable building block.

The top-level controller is small and simple enough that it could easily be implemented as a GRN. Alternately, a brain could presumably implement it easily as well (which is relevant to examples of brain control of morphogenesis [34]). For simplicity, we have merely left it as software in our simulations.

Simulation results

We show three simulations of the system with different $N$. In the first simulation, our goal is two peaks: i.e., a morphogen profile of LHLH. We start out (Table 1) with $\lambda_{RD}=7.2 \times 10^{-7}$, which yields four peaks rather than the desired two. The top-level controller then directs seven iterations of counting, noting that there are too many peaks, and increasing $\lambda_{RD}$ by 10% on each iteration, eventually giving the desired two peaks. The second simulation (Table 2) starts with the same initial conditions but has a goal of $N=5$ rather than $N=2$. This time, the controller goes through 4 iterations of decreasing $\lambda_{RD}$, eventually creating the desired extra peak.

The third simulation (Table 3) shows an interesting quirk of some RD patterns. As noted above, a single field length $L$ can often support various numbers of peaks; it is a dynamic system with multiple stable points, and it is often difficult to know which stable point the system will travel to. We thus see iteration #1 setting $\lambda_{RD}=5.4 \times 10^{-7}$, which could have given us five peaks but instead gave six. Iteration #3 hopped over five peaks directly to four. Eventually, though, the controller keeps adjusting $\lambda_{RD}$ until it successfully reaches the goal.

Varying segment lengths

We have shown that varying $\lambda_{RD}$ allows us to control the number of RD peaks we create. Once we have achieved the desired peak count, we can then vary $\lambda_{RD}$ locally to further control pattern shape. Figure 2a shows a pattern with 100 cells that was targeted to a three-peak pattern. The peaks are originally at cells 20, 60 and 100 (blue graph). We have then added one more pass to our top-level controller, after it has converged on the correct number of peaks. In that extra pass, the GRN is slightly modified so that any cell expressing *S0H* increases its $\lambda_{RD}$ by 60%

(Figure 2a, orange graph). Figure 2b is quite similar, except in this case we have modified the GRN so that any cell expressing either *S0H* or *S1L* increases its $\lambda_{RD}$ by 60%. In each case, the appropriate segment(s) of the RD pattern increase in length at the expense of their immediate neighbors.

This capability is our digital equivalent of what Green[12] calls "RD acting downstream of PI." In his version, PI first lays down a coordinate system and then RD uses it to affect $\lambda_{RD}$. In ours, our cellular automaton first lays down one or more coordinate systems and we can then stretch or shrink any subset of them fairly arbitrarily.

GRN details and limits of robustness

Here is the detailed GRN that implements our cellular automaton:

```
Pre1L = Pre1L | nothing          | [S1L&(M<Schm1)& !done]
Pre1H = Pre1H | [S1L&(M>Schm1)& !done] | [S1H&(M>Schm0)& !done]
Pre2L = Pre1L | [S1H&(M<Schm0)& !done] | [S2L&(M<Schm1)& !done]
Pre2H = Pre2H | [S2L&(M>Schm1)& !done] | [S2H&(M>Schm0)& !done]
…
done = Pre1H | Pre1L | Pre2H | Pre2L | …
S1L=Pre1L
S1H=Pre1H
S2L=Pre2L
S2H=Pre2H
```

Cells communicate with each other with the *S\** signals. Any cell participates in the computation wave by waiting to receive an *S\** signal, then deciding which *S\** signal to relay onwards. The *Pre\** and *done* signals implement the wavefront concept. Any cell, once it receives an *S\** signal, makes its decision by driving one of the *Pre\** signals. These signals stay within the cell and are purely for internal bookkeeping. Once a cell drives any of its *Pre\** signals high, its *done* (which also remains in the cell) also goes high. This then feeds back to the first set of equations and serves to cut off the *Pre\** signals from looking at any incoming *S\** signal any longer. At this point, the self-loop in each *Pre\** equation takes over, so that whichever *Pre\** signal is asserted will stay asserted.

Finally, the appropriate *S\** signal gets driven out of the cell, and stays asserted until a *reset* comes in and breaks the *Pre\** self-loops. *Reset* is a global (i.e., widely-diffusing) signal that operates by substantially increasing the degradation rate of the *Pre\** signals (e.g., by adding a degradation tag). This breaks the self-loop, and thus turns off all *Pre\** and then *done* signals in each cell.

Given that each *S\** signal is merely a buffer of its corresponding *Pre\** signal, why bother with the extra signals? In our previous rough description of the GRN, we used two versions of each *S\** signal, one "_in" and one "_out" at each cell. In reality, cells do not have unidirectional signal ports, and this signal duplication essentially fixes that problem. The self-loop on each *Pre\** signal implements our memory of the decision a cell takes; if we tried to put that self-loop on the *S\** signal instead, then each cell would latch the incoming *S\** signal before making a decision of its own.

We implement the logic equations, as is often done, as Hill functions. The *Pre\** gates are the most complex: e.g., for *Pre0H*:

$$Pre0H = k_v \frac{\left(\frac{Pre0H}{.5}\right)^3}{1+\left(\frac{Pre0H}{.5}\right)^3} + \frac{(S0L/.3)^3}{1+(S0L/.3)^3} \cdot \frac{(A/.3)^3}{1+(A/.3)^3} \cdot \frac{1}{1+(done/.5)^3} + \frac{(S0H/.3)^3}{1+(S0H/.3)^3} \cdot \frac{(A/.1)^3}{1+(A/.1)^3} \cdot \frac{1}{1+(done/.5)^3}.$$

The *done* gates are implemented as $done = k_v \frac{\left(\frac{\sum Pre*}{.5}\right)^3}{1+\left(\frac{\sum Pre*}{.5}\right)^3}.$

Finally, the $S*$ gates are (e.g., for $S0H$) $S0H = k_v \frac{\left(\frac{Pre0H}{.5}\right)^3}{1+\left(\frac{Pre0H}{.5}\right)^3}.$

Robust operation of the computation wave places some constraints on system parameters. For example, the *done* and *Pre\** signals must be species that do not travel between cells, thus allowing each cell to decide for itself when the computation wave has reached it. The $S*$ signals are meant to travel by diffusion to their nearest neighbors. A molecule $S$ generated at a constant rate $G_S$ moles/s at $x_0$ and diffusing freely will have its concentration given by $G(x) = \frac{G_S}{K_{D,S}} e^{-\lambda_S |x-x_0|}$, where $K_{D,S}$ is the decay rate for $S$, $\lambda_S = \sqrt{\frac{D_S}{K_{D,S}}}$ and $D_S$ is the diffusion rate for $S$. Clearly $\lambda_S$ must be large enough for the signal to reach its nearest-neighbor cells, which is hopefully easy. However, it must also be small enough that the $S*$ signals do not travel further than one half cycle of the pattern. So, e.g., $S1H$ will first be generated at the cell $C_0$ where $S1L$ is seen and $[A]>.3$; it will correctly be generated at cells further to the right until we reach a cell $C_1$ where $[A]<.1$. At that point, $S2L$ will correctly be generated, and regenerated at successive cells until we reach a cell $C_2$ where $[A]>.3$ again. But what if $S1H$ (traveling by diffusion from cell $C_1$) reaches cell $C_2$ before $S2L$ (being regenerated at each cell from $C_1$ to $C_2$) does? In that case, cell $C_2$ will incorrectly see $S1H$ as the incoming wavefront, and will express $Pre1H$ immediately, and will be locked into that decision before it sees $S2L$ and tries to express $S2H$. As a result, the count will be too low by one.

How do we avoid this issue? Diffusion is an order(distance squared) process and thus slow over long distances. Our trick then simply to be sure that the half-pattern-length distance is always more than just one or two cells. This imposes a minimum size on $\lambda_{RD}$.

Details of the RD pattern

There are countless varieties of RD equations in the literature. Most would work equally well in our system, but we had to choose one. We used a very simple RD pattern taken from [33], where $\frac{\partial A}{\partial t} = g_A G(A,I) - k_{D,A} A + D_A \frac{\partial^2 A}{\partial x^2}$, $\frac{\partial I}{\partial t} = g_I G(A,I) - k_{D,I} I + D_I \frac{\partial^2 I}{\partial x^2}$ and $G(A,I) = \frac{1}{1+(I/A)^h}.$

In [33] (and indeed in much of the RD literature), the reaction and diffusion can occur anywhere in a homogeneous fixed-length field. Instead, we have chosen to implement our 1D field as cells interconnected by gap junctions (GJs). Each cell holds one GRN in our cellular automaton. It would be perfectly reasonable for the RD activator $A$ and inhibitor $I$ to be small molecules that could diffuse through a cell membrane and then interact anywhere in the field. Instead, we chose to interconnect the field of cells with GJs, which we assume form the conduit that $A$ and $I$ diffuse through. We made this choice primarily for simplicity (our existing simulation framework supports it well), and it is not central to our work. However, there is substantial evidence[35-37] in fish models of RD molecules traveling through GJs, so this seemed to be a reasonable choice.

The feedback mechanism in our system operates by repeatedly adjusting $\lambda_{RD}$. Since $\lambda_{RD} = \sqrt{\frac{D_A}{K_{D,A}}}$ (where $D_A$ is the diffusion constant of $A$ in m$^2$/sec and $K_{D,A}$ is the degradation constant of $A$ in sec$^{-1}$), this can be done by either adjusting $D_A$ or $K_{D,A}$. Nature has access to multiple means of controlling both of these[22]; degradation tags, competing reactions to bind a morphogen, and even lipid modification to affect diffusivity. We choose to alter $D_A$ by changing GJ density. Altering the density of GJs between cells changes the effective diffusion rate of molecules passing through those GJs. While there is as yet no well understood mechanism in nature for controlling GJ density, there is substantial evidence that such a mechanism must exist. The mammalian heart conducts action potentials between cardiac cells primarily via GJs. If the density of GJ interconnect varies outside of a prescribed range, cardiac conduction fails[38]; and cardiac operation is nothing if not extremely reliable. Hence such mechanisms must exist, though again we do not understand them. Once more, we choose to use this mechanism in our simulations for compatibility with an existing simulator, and it is not central to our results.

Choosing a feedback measure

Robustness is one of the great challenges in morphogenesis [22], and many strategies have evolved to achieve it. Many of them use negative feedback, which is an extremely common motif in nature [28]. For example, mice use RD to pattern the toes on their feet [23]. To then prevent large embryos from having extra toes, mice use fibroblast growth factor (FGF), which affects embryo growth, to also increase $\lambda_{RD}$. If FGF were the only factor affecting embryo growth, this strategy would be completely robust. However, in addition to the inherent unreliability in RD, any variations in embryo size from sources other than FGF are not controlled for, thus occasionally resulting in four- or six-toed mice.

In error-correcting control systems, you get what you measure. FGF concentration is serving as a proxy (albeit an imperfect one) for the field length $L$, and being used to control for the fact that increasing $L$ would normally increase the number of RD pattern replications. In other words, the proxy for $L$ is being fed *forward* to control $\lambda_{RD}$. However, arguably the most appropriate goal is to preserve the number of toes – and the most reliable way to do that is not to measure $L$ at all, but to directly measure the number of toes as a feedback mechanism.

Expansion-repression basically generates $A$ at one end of the field at some rate $G_A$, measures [$A$] at the other end, and increases(decreases) $\lambda_{RD}$ when the distal [$A$] is too small(large). It is thus using the distal [$A$] as a proxy for $L$. If $L$ doubles, then expansion-repression will double $\lambda_{RD}$ (e.g., by quadrupling $D_A$). The combination of doubling $L$ and quadrupling $D_A$ can easily be shown to exactly restore the original [$A$] profile. Since [$A$] affects $\lambda_{RD}$, which then affects [$A$], this is indeed negative feedback.

If, instead of doubling $L$, we double $G_A$, then the distal [$A$] will originally double. It is also easy to show that if we then double both $D_A$ and $K_{D,A}$, then the profile of [$A$] will be restored. However, in both these cases, if we restore the distal [$A$] by any other combination of changing $D_A$ and $K_{D,A}$, then [$A$] at the source will change, as well as the exact profile. The issue is that we are attempting to preserve an exact profile shape but using [$A$] at a single point to serve as a proxy for that profile.

Our top-level loop, measuring the number of peaks and altering $\lambda_{RD}$ accordingly, is clearly a negative-feedback system. Specifically, our feedback variable is the number of pattern peaks, which is exactly the final variable most important to control. Thus, as long as our feedback system itself is operational, we will be immune from changes in other, non-feedback variables.

This is particularly important since RD patterns are not fully predictable. As noted above, while a pattern with characteristic length $\lambda_{RD}$ will be stable on a field of length $L$, it may also be stable on any field of length longer than $L$. Similarly, increasing $\lambda_{RD}$ such that (as noted above) a six-peak pattern is no longer viable may lead instead to a four-peak pattern, even though five peaks would be feasible. A field can thus often stably sustain a choice of multiple RD peak counts. Each choice has its own stability region around it, where unstable initializations will flow to that particular stable point. The stability regions are often difficult to predict.

The basic building blocks of biology are almost always noisy and difficult to predict [39]. Building systems that nonetheless work reliably is thus often difficult, and our case is no exception. While our basic RD patterns can be difficult to use, the most effective feedback system – one where we close the loop by directly monitoring the variable we care most about – is quite effective. This is exactly the system we have built.

## Discussion

We have described a cellular automaton, working within a simple negative-feedback controller, that takes an existing field of cells, subdivides it into smaller pieces and lays down a coordinate system in each piece. The heart of the system is the ability to count the number of peaks in a morphogen pattern. This is similar to the classic majority-detection problem for a cellular automaton, where you must look at a field of cells that are either 0 or 1, and determine if there are more 0s or more 1s. While the problem may seem simple, it is not at all so [40, 41]. A classical cellular automaton, in this problem, is defined as working with local information without the benefit of knowing head vs. tail – and yet is attempting to solve a global-scale problem.

But counting is a simple operation for most human six-year-olds. We of course have not solved the well-studied majority-detection problem. Instead, we take advantage of having a prior process break symmetry and give us a known head and tail, thus enabling a global unidirectional sweep to simply count. Thus, our model is a contribution to the classic problem of leveraging large-scale morphogenetic order from molecular symmetry breaking [42-45].

Arguably, subdividing an existing field is not a universal situation, since many kinds of embryos are growing at the same time as cells are differentiating. It is not uncommon, though. Mammalian and avian blastoderms can divide in two to create identical twins; each of the two new embryos then reforms itself, differentiating anew to alter each of the existing embryos.

Planarian morphallaxis is another interesting case. When an adult planarian is cut into fragments, each fragment can regrow into a full new worm. However, since a fragment may be missing a mouth or indeed an entire digestive system, the fragment cannot increase its mass until it has the capability of eating. It thus undergoes morphallaxis [46, 47], where the fragment reforms itself into a fully formed but small-scale planarian, and then eats and grows.

A human body is about 30,000 cells tall. PI in its basic form is not capable of determining the correct position of every cell – that would require reading the morphogen gradient to four significant digits, which is not biologically realistic. A more plausible strategy is to divide and conquer, where the body first divides into (e.g.) organs, which then form and differentiate independently from each other, each using its own smaller coordinate system. More complex organisms might create their form using even more levels of hierarchy. The task of forming a foot might involve subdivision into five smaller pieces, each with its own coordinate system, and then using a common "toe routine" to further develop each identical piece.

Our closed-loop RD system can do this easily, partitioning the partially-grown field of cells into smaller pieces, each with its own coordinate system. We may further want different toes to take up different amounts of space, with a big toe typically being wider than the others, which would use our capability for unequal subdivision.

Our computation-wavefront implementation has striking similarities to how neurons operate. While action potentials typically travel unidirectionally along a neuron, *antidromic* (i.e., "reverse-direction") propagation is also possible. In fact, clinical EMG studies routinely employ antidromal propagation in a *F-wave* test to quantify nerve-conduction velocity [48].

The reason that action potentials do not spontaneously reverse direction while traveling down a neuron is that the segment of an axon "behind" the AP is still refractory, and thus cannot retrigger with a reverse-direction wave. This is essentially how our computation wavefront works; while our signaling molecules can (and do) travel in the wrong direction, upstream cells are essentially refractory by virtue of having already triggered, and thus ignore the undesired signal.

We can make an interesting comparison to Alamia [49, 50], where they model a visual cortex using predictive coding as two interconnected modules. The lower-level module senses its environment and produces an error signal comparing that environment to a prediction; the upper-level module looks at the error signal and updates the internal model and gives the new model back to the lower-level model. The system oscillates, duplicating known characteristics of alpha oscillations.

It is well known [28, 51] that negative-feedback systems can oscillate if, at the frequency where their loop delay is 180º, their loop gain is greater than one. Indeed, we ensured that we increment $\lambda_{RD}$ by fairly small amounts to minimize our loop gain and help ensure convergence. What about the loop delay? Two interconnected analog systems have delay given by their circuit structure. However, we have again taken advantage of asymmetry to effectively make our delay constant and minimal. The computation wave starts at the organism's tail; the top-level controller lives in the head and can wait until wave is finished before starting the next iteration. Our analog delay is thus out of the picture. Alamia further builds a multi-level model of a visual cortex [49] and notes the resulting traveling waves; while our model is only two levels, a larger morphogenesis model would likely also be multi-level and may show similar results.

This work is purely *in silico*; we have yet no evidence that this particular cellular automaton exists in nature. However, it seems fairly clear that *some* sort of negative-feedback system must exist. The ability of a mammalian embryo to successfully recover from disturbances as varied as being split in two (e.g., for identical twins) and transient mRNA interference to the early embryo [52] would be difficult to explain otherwise. Regardless of whether evolution found exactly this scheme, it can now be used in synthetic biology approaches to engineer novel patterning systems [53-56].

Even outside of RD, a cellular automaton using computation wavefronts to process information directionally (e.g., from tail to head) seems useful. Consider a simple PI system, with a GRN in each cell implementing the ubiquitous French flag decoding. At the boundary between, say, the blue and white stripes, noise will blur the decision. Instead of a consecutive group of cells ideally interpret their morphogen levels as BBBBWWWW (i.e., a sharp boundary between blue and white in the middle of the field), it may be, e.g., BBWBBWWW; the boundary cells come to independent and fairly unpredictable decisions. One pass with a computation wavefront from left to right and finding the first W could make this boundary quite sharp, in this case giving BBWWWWWW.

Lander[22] has observed that robustly creating sharp borders in morphogenesis is an inherently hard problem – the positive feedback that helps make borders sharp also tends to amplify noise, resulting in an unpredictable location (where our French-flag example declared the boundary to be a the first cell that, given the noise, declared itself to be white). A digital system such as ours can inherently create very sharp boundaries; a *multipass* digital system might be able to do so quite robustly. We believe that the ability to implement unidirectional computation with wavefronts is quite powerful and may have many applications.

**Conclusions**

We have demonstrated a closed-loop negative-feedback machine to control morphogenesis, as a contribution to the efforts to understand morphogenesis as a target-directed process [11]. Its goal is to lay down *N* copies (for a reasonably arbitrary *N*) of a simple RD pattern; e.g., to be used as repeats of a coordinate system. It achieves this goal with a closed-loop negative-feedback controller that

- employs a cellular automaton to counts peaks, and thus count the current number of pattern repetitions. It uses *computation wavefronts*, a powerful concept in cellular computing that takes advantage of existing asymmetry .
- Compares the current number of peaks to its goal *N*.
- adjusts the RD pattern-length parameter $\lambda_{RD}$ so as to move the pattern towards the goal.

We have further described a way to controllably make a subset of the patterns larger or smaller than the others, so as to subdivide a field into unequal subfields in a repeatable manner.

The circuit described here enables flexible actions under a range of circumstances to reach a specific large-scale goal state (a multi-cellular morphogenetic prepattern) which belongs to the collective and not the individuals [57, 58]. Such capacity has been proposed as a definition of intelligence [10], for the field of basal cognition [59-63]. Thus, the above circuit and analysis not only supports a way of viewing morphogenetic processes as a set of specific computational tasks, but also suggests an architecture for incorporating a simple kind of intelligence into novel biological constructs [64-70]. Future work will investigate the presence of these dynamics in vivo, as well as use the insights revealed by this modeling process to create novel patterning systems for synthetic biorobotics, regenerative medicine, and tissue bioengineering.

**Acknowledgements**

M.L. gratefully acknowledges support of the John Templeton Foundation (62212).

| Iteration | $\lambda_{RD}$ | n_peaks | $N$ | action |
|---|---|---|---|---|
| 0 | $7.2\times10^{-7}$ | 4 | 2 | init |
| 1 | $7.9\times10^{-7}$ | 4 | 2 | increase |
| 2 | $9.6\times10^{-7}$ | 4 | 2 | increase |
| 3 | $1.1\times10^{-6}$ | 4 | 2 | increase |
| 4 | $1.2\times10^{-6}$ | 3 | 2 | increase |
| 5 | $1.3\times10^{-6}$ | 3 | 2 | increase |
| 6 | $1.4\times10^{-6}$ | 3 | 2 | increase |
| 7 | $1.5\times10^{-6}$ | 2 | 2 | increase |

Table 1. Summary of simulation #1. Each row is one top-level-algorithm iteration. N_peaks is the current number of peaks found, and N is the target. Action describes the action of the top-level algorithm: whether it is increasing vs. decreasing $\lambda_{RD}$.

| Iteration | $\lambda_{RD}$ | n_peaks | $N$ | action |
|---|---|---|---|---|
| 0 | $7.2\times10^{-7}$ | 4 | 5 | init |
| 1 | $6.5\times10^{-7}$ | 4 | 5 | decrease |
| 2 | $5.9\times10^{-7}$ | 4 | 5 | decrease |
| 3 | $5.4\times10^{-6}$ | 5 | 5 | decrease |

Table 2. Summary of simulation #2

| Iteration | $\lambda_{RD}$ | n_peaks | $N$ | action |
|---|---|---|---|---|
| 0 | $4.9\times10^{-7}$ | 6 | 5 | init |
| 1 | $5.4\times10^{-7}$ | 6 | 5 | increase |
| 2 | $5.9\times10^{-7}$ | 6 | 5 | increase |
| 3 | $6.5\times10^{-7}$ | 4 | 5 | increase |
| 4 | $5.9\times10^{-7}$ | 4 | 5 | decrease |
| 5 | $5.4\times10^{-7}$ | 5 | 5 | decrease |

Table 3. Summary of simulation #3

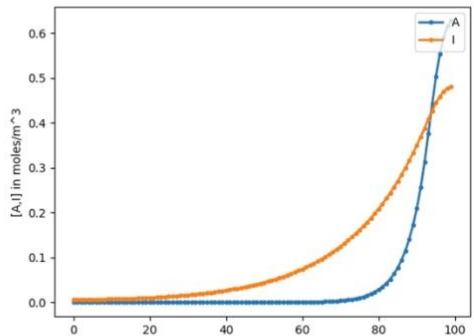 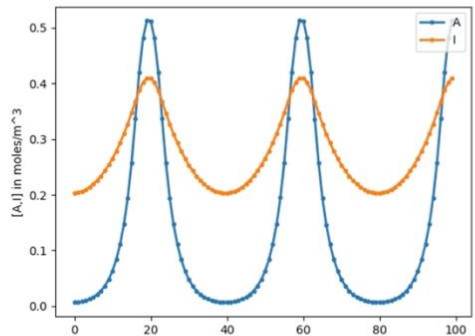

(a) 1 peak pattern (LH)

(b) 3 peak pattern (LHLHLH)

**Figure 1. RD pattern profiles in 1D**
(a) shows a simple one-peak pattern, otherwise called LH (low [A] on the left and high [A] on the right). (b) shows a three-peak pattern (LHLHLH). Note that the inhibitor *I* spreads wider than *A* due to its higher diffusion coefficient.

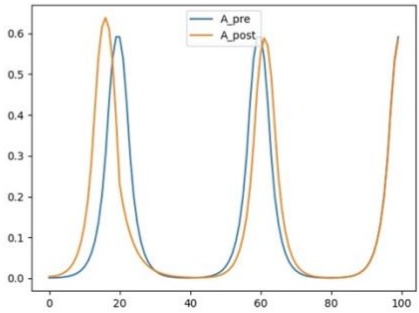 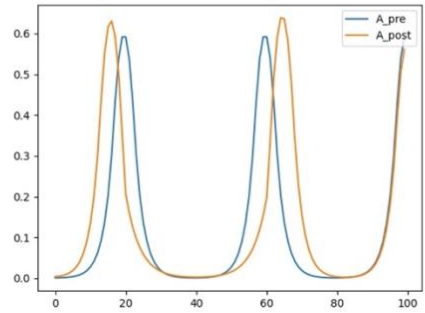

(a) Cells expressing *S0H* increase $\lambda_{RD}$

(b) Cells expressing *S0H* or *S1L* increase $\lambda_{RD}$

**Figure 2. Skewing $\lambda_{RD}$ digitally**
These graphs are the result of an a post-process that uses the "bread-crumb" signals *S\** created by the cellular automaton. The blue lines are the evenly-spaced signals before skewing; the orange shows the results after intentional skewing.

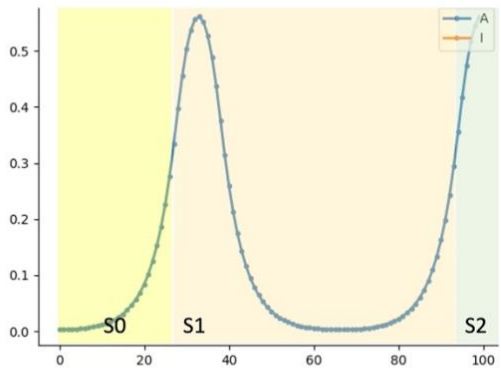
(a) S0, S1 and S2 by location

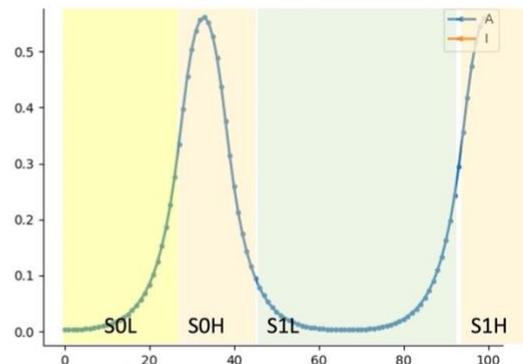
(b) Schmidt-trigger signals by location

**Figure 3. Which signals are expressed by which cells in the automaton**
As an aid to understanding how the cellular automaton works, we graphically show which cells express each of the S* signals. (a) is for the original algorithm, and (b) is for the improved algorithm using Schmidt triggers.

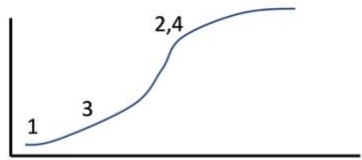

In this graph, the x axis is cell number and the y axis is [A]. In step 1, the left side of the field (x=0) is seeded with SOL. Successive cells from left to right replicate this signal. In step 2, the signal reaches a cell where [A] is high enough that we now generate SOH. This signal travels to the right as desired (not shown), but also to the left. In step 3, SOH reaches a cell where [A] is low enough to generate S1L. This then diffuses as usual, eventually reaching step 4, where the same cell that originally saw SOL and correctly generated SOH now sees S1L and mistakenly generates S1H. The result is a peak getting double counted.

**Figure 4. The problem with loops**

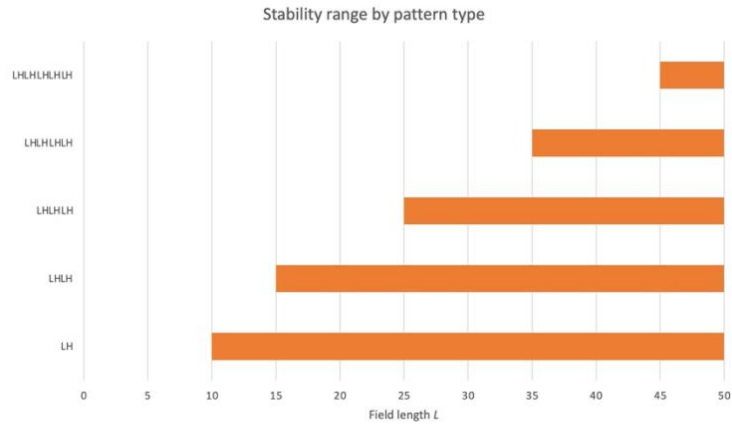

**Figure 5. Pattern viability at various L values**
This figure illustrates how most field lengths can support more than one RD pattern. The LH pattern is viable at all field lengths larger than $\lambda_{RD}$. At field lengths $L$ in the range [15,20], both LH and LHLH are viable. At longer $L$, still more shapes are viable.